\documentclass[12pt]{iopart}
\usepackage{iopams}
\usepackage{graphicx}
\expandafter\let\csname equation*\endcsname\relax
\expandafter\let\csname endequation*\endcsname\relax
\usepackage{amsmath}
\usepackage{amssymb}
\usepackage{todonotes}
\usepackage{hyperref}
\usepackage{color}
\usepackage{xcolor}
\usepackage{tikz}
\usetikzlibrary{calc}
\usetikzlibrary{arrows.meta}
\usetikzlibrary{shapes.geometric}

\begin{document}
\title[The quantum harmonic oscillator on a circle]{The quantum harmonic oscillator on a circle -- fragmentation of the algebraic method}
\author{Daniel Burgarth }
\address{Department Physik, Friedrich-Alexander-Universität Erlangen-Nürnberg, Staudtstraße 7, 91058 Erlangen, Germany}
\author{Paolo Facchi}
\address{Dipartimento di Fisica, Universit\`a di Bari, I-70126 Bari, Italy}
\address{INFN, Sezione di Bari, I-70126 Bari, Italy}

\begin{abstract}
    A quantum particle on a circle in a quadratic potential exhibits a spectrum that is not harmonic, despite having all algebraic properties of the quantum harmonic oscillator. This raises the question where the usual algebraic argument ---implying integer gaps--- fails. The answer is illuminating and covers a  surprisingly rich range of physical phenomena for such a simple model.
\end{abstract}
\section{Introduction}
The  harmonic oscillator is the workhorse of theoretical physics. In suitable units $\omega=m=\hbar=1$ its quantum Hamiltonian is given by \begin{equation}\label{ham}\frac{1}{2}\left(p^2+q^2\right)\end{equation} where $q$ and $p$ are the canonical position and momentum operators respectively.

The beauty of the harmonic oscillator stems from the fact that much of it can be understood from a purely algebraic perspective without differential equations or functional analysis. This, and its importance in all areas of quantum mechanics and beyond, makes it an attractive subject of undergraduate teaching. One usually defines creation $a^\dagger= (q-\textrm{i}p)/\sqrt{2}$ and annihilation $a=(q+\textrm{i}p)/\sqrt{2}$ operators. From the canonical commutation relations $[q,p]=\textrm{i}$ follows $[a,a^\dagger]=1$, and the Hamiltonian is factorized as 
\begin{equation}
  a^\dagger a+\frac{1}{2}.  
\end{equation}
 Using only the positivity of the Hamiltonian and the commutation relation of $a$ and $a^\dag$, one then finds the usual harmonic spectrum $n+\frac{1}{2}$ where $n\in \mathbb{N}$. The key features are integer distances between energies and a ground state with non-vanishing energy, which can be intuitively understood with Heisenberg's uncertainty principle.

The purpose of the present article is to study what can go wrong with the algebraic proof when one slightly deviates from the well-trodden path  and changes the Hilbert space. Rather than considering a particle on a line $\mathbb{R}$, we shall consider a particle on a circle $\mathbb{S}_r$ of radius $r>0$. While we keep the rest the same, we shall see that this changes (almost) \emph{everything.}  

The structure of the paper is as follows. In Section \ref{sec2} we set up the canonically conjugated position and momentum operators on the circle and discuss their commutator. In Section~\ref{sec:fullcircle} we discuss the harmonic oscillator and find its spectrum and asymptotics. We define the creation/annihilation operators in Section~\ref{creation} and factor the Hamiltonian in normal and anti-normal order. Then, we discuss the key question of this article in Section~\ref{algebraic}, namely, what happens to the algebraic argument. Section~\ref{appA} shows that the factorization leads quite naturally to an infinity of contestants for the harmonic oscillator, and we discuss their spectrum in Section~\ref{appB}. We conclude in Section~\ref{conclusions}.

\section{Relevant Literature}

The central goal of our work is to investigate the fragmentation of the standard algebraic construction for the harmonic oscillator, as presented in  textbooks~\cite{LandauLifshitzQM,SakuraiQM,CohenTannoudjiQM}. That there might be issues for position and momentum on the circle was already hinted at in \cite[11.2]{hall_quantum_2013} although not explored in detail. More generally, the failure of Lie algebra representations of unbounded operators to exponentiate to group representations is a well studied area in  mathematical physics, see for instance~\cite{schmudgen_invitation_2020}. Related issues have also been discussed by the present authors in the context of orbital angular momentum on the torus~\cite{wrongangular}. 

The functional-analytic tools required for our analysis can be found in  standard references, such as~\cite{Teschl,reed_methods_2005}. A central role is played by boundary conditions and self-adjoint extensions, a rich area of quantum theory. See, for instance, Refs.~\cite{AsoreyIbortMarmo2005,AsoreyIbortMarmo2015,deOliveira2009,FacchiGarneroLigabo2018}, and, in the context of form domains and singular perturbations,  Refs.~\cite{SimonQuadraticForms,BehrndtHassiDeSnoo2020,ExnerGrosse1999,Albeverio2005}. Concrete realizations have been analyzed  on the segment in Ref.~\cite{AngeloneFacchiMarmo2022}, and on the circle in Refs.~\cite{FulopTsutsuiCheonJPSJ,angelone}. Pedagogical issues arising from unbounded operators, together with a nice collection of seeming paradoxes related to boundary conditions, are discussed in~\cite{BonneauAJP}, while the interplay between Galerkin approximations and self-adjoint extensions has been explored in~\cite{wrongbox}. The novelty of our approach lies in connecting the partial factorization of the quadratic Hamiltonian of the harmonic oscillator to a $U(1)$ family of extensions and in analyzing the resulting algebraic features.

Our work is also connected to developments in deformed oscillator algebras and phase-related quantization schemes. In particular, the $q$- and $f$-oscillator frameworks show how modified commutation relations naturally lead to nonlinear spectra and altered ladder structures~\cite{Manko1993,Manko1997,Rego-Monteiro2001,Kastrup2007,Susskind1964}.

Further connections can be drawn with spectral-theoretic studies of nonstandard oscillator models, including the complex rotated harmonic oscillator, where the spectrum remains discrete but the basis properties become subtle, and complex anharmonic oscillators, where spectral projections may fail to form a Riesz basis. These works reinforce the broader message that oscillator-like operators can retain formal algebraic features while exhibiting significantly different spectral and functional-analytic behavior~\cite{Davies1999,DaviesKuijlaars2004,BalmasedaKrejcirikPerezPardo2025,MityaginSiegl2026}.

From a physical perspective, an interesting debate about the role of compact and non-compact manifolds has emerged in the context of superconducting systems~\cite{Devoret2021,LeColeStace2020}. The harmonic oscillator on a segment with boundary conditions also play a crucial role in the quantum Hall effect,  as discussed in Ref.~\cite{AngeloneAsoreyFacchiLonigroMartinez2023}. Finally, compact manifolds  play an important role in several areas of high-energy physics, including string theory and gauge theory, where oscillator algebras often enter in particular algebraic constructions; see, for example, Refs.~\cite{CremadesIbanezMarchesano2004,Troost2000,vanBaal1984}.

\section{Hilbert space on the circle and canonically conjugated operators}\label{sec2}
We will now set up canonically conjugated operators on the Hilbert space $L^2(\mathbb{S}_r)$ of the square integrable functions (wavefunctions) on a circle (see also \cite[Section 12.2]{hall_quantum_2013}). 
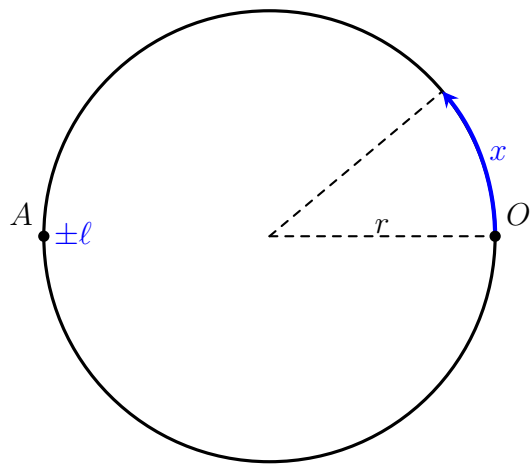
\begin{figure}
    \centering
\begin{tikzpicture}[scale=3, line cap=round, line join=round]

% Parameters
\def\r{1}            % radius
\def\angO{0}         % angle of point O (degrees)
\def\angArc{40}      % arc length angle (degrees, counterclockwise)
\def\labelR{1.08}    % radius for arc label placement

% Center
\coordinate (C) at (0,0);

% Points on the circle
\coordinate (O) at ({\r*cos(\angO)},{\r*sin(\angO)});
\coordinate (A) at ({\r*cos(\angO+180)},{\r*sin(\angO+180)});
\coordinate (E) at ({\r*cos(\angO+\angArc)},{\r*sin(\angO+\angArc)});

% Circle
\draw[black, very thick] (C) circle (\r);

% Radius to O (dashed)
\draw[black, dashed, thick] (C) -- (O)
  node[midway, above, fill=white, inner sep=1pt] {$r$};

% Radius to arc endpoint (dashed)
\draw[black, dashed, thick] (C) -- (E);

% Blue counterclockwise arc with strong arrow tip
\draw[blue, ultra thick,
      -{Stealth[length=6pt, width=6pt]}]
  (C) ++(\angO:\r)
  arc[start angle=\angO, end angle=\angO+\angArc, radius=\r];

% Arc label x (placed off the arc)
\node[blue] at ({\labelR*cos(\angO+0.5*\angArc)},
                {\labelR*sin(\angO+0.5*\angArc)}) {$x$};
\node[blue] at (-0.87,0.01) {$\pm\ell$};
%\node[blue] at (-0.87,-0.1) {$ -\ell$};
% Mark and label points
\fill (O) circle (0.025) node[above right] {$O$};
\fill (A) circle (0.025) node[above left] {$A$};

\end{tikzpicture}
    \caption{Geometry of our setup and the choice of position $x$}
    \label{fig:circle}
\end{figure}
The position operator acts as a real multiplication operator by the signed arc-length from a fixed point $O$ on the circle of circumference $2\ell = 2\pi r$  
\begin{equation}
    (q \psi)(x)=x \psi(x).
    \label{eq:qdef}
\end{equation}
However, because the value of $x$ is between $[-\ell,\ell)$, with the endpoints $\pm\ell$ both corresponding to the antipodal point $A$ (see Fig.~\ref{fig:circle}), this is now a \emph{bounded} operator with norm $\|q\|=\ell$. It is therefore self-adjoint and defined on the whole Hilbert space, $D(q)=L^2(\mathbb{S}_r)$. The momentum operator acts as differentiation,
\begin{equation}
    (p \psi)(x)=-\textrm{i}\psi'(x),
    \label{eq:pdef}
\end{equation}
with domain of self-adjointness $D(p)=H^1(\mathbb{S}_r)$, where
\begin{equation}
    H^1(\mathbb{S}_r) = \left\{\psi \in L^2(\mathbb{S}_r) \, : \, \psi' \in L^2(\mathbb{S}_r) \right\}
\end{equation}
is the first Sobolev space of wavefunctions on the circle whose (weak) derivative is still a wavefunction (see also \ref{appC}). This is an unbounded operator with pure point spectrum $\mathrm{spec}(p)=\frac{\pi}{\ell} \mathbb{Z}=\frac{1}{r} \mathbb{Z}$, with eigenfunctions forming the Fourier basis $\{e^{i n\frac{x}{r}}/\sqrt{{2\pi r}}\,:\, n\in\mathbb{Z}\}$.

The canonical commutation relation 
\begin{equation}
 [q,p]=\textrm{i}   
\end{equation}
holds on a dense domain. Indeed, the commutator is defined on the domain $D([q,p])= D(q p)\cap D( p q)$, with 
\begin{eqnarray}
    D(q p) &=& \left\{\psi\in D(p) \,:\, p\psi \in D(q)\right\}= \left\{\psi\in H^1(\mathbb{S}_r) \,:\, -i\psi' \in L^2(\mathbb{S}_r)\right\}
    \nonumber\\
    &=& H^1(\mathbb{S}_r) = D(p),
\end{eqnarray}
so that
\begin{eqnarray}
    \fl \quad D(q p)\cap D(p q) = \left\{\psi\in D(p) \,:\, q\psi \in D(p)\right\}= \left\{\psi\in H^1(\mathbb{S}_r) \,:\, x\psi(x) \in H^1(\mathbb{S}_r)\right\}.
\end{eqnarray}
Now $x$ is bounded and $C^{\infty}(\mathbb{S}_r\setminus \{A\})$ and thus $\psi\in H^1(\mathbb{S}_r)$ implies that $x\psi(x)$ is in $H^1(\mathbb{S}_r\setminus \{A\})$. Moreover, $H^1(\mathbb{S}_r)\subset C(\mathbb{S}_r)$ so that $x \psi(x)$ should be continuous at $A$, i.e.\ $\psi(\pi r)= \psi(-\pi r)=0$. Therefore, the domain of the commutator reads
\begin{equation}
     D([q,p]) = \left\{\psi\in H^1(\mathbb{S}_r) \,:\, \psi(\pi r)=0 \right\},
\end{equation}
and is dense in $L^2(\mathbb{S}_r)$. Moreover,
for any $\psi \in D([q,p])$ one has that
\begin{equation}
[q,p]\psi(x) = - i x \frac{d}{dx}\psi(x) + i  \frac{d}{dx} \bigl(x \psi(x)\bigr) = i \psi(x).
\end{equation}

Notwithstanding the fact that $q$ and $p$ are canonical conjugate variable, exactly as on the line, the Weyl relations $e^{itq}e^{isp}=e^{isp}e^{itq}e^{-i t s}$ do \emph{not} hold \cite[Example 14.5]{hall_quantum_2013}. This means that the usual algebraic arguments appear to remain the same, yet the physics is very different. This holds also for the harmonic oscillator built from such modified operators.

\section{Going full circle with the harmonic oscillator\label{sec:fullcircle}}
Now  let us consider the harmonic oscillator on the circle of radius $r$. The Hilbert space is $L^2(\mathbb{S}_r)$. The Hamiltonian is 
\begin{equation}
   H= \frac{1}{2}\left(p^2+q^2\right)
   \label{eq:hoHamdef}
\end{equation}
with $q$ and $p$ the conjugated operators in~\eqref{eq:qdef} and~\eqref{eq:pdef}, which are self-adjoint on $D(q)=L^2(\mathbb{S}_r)$ and $D(p)=H^1(\mathbb{S}_r)$. Since $q$ is bounded, $q^2$ is also bounded and defined, and self-adjoint, on the whole Hilbert space $D(q^2)=L^2(\mathbb{S}_r)$. On the other hand the domain of self-adjointness of $p^2$ is
\begin{equation}
    D(p^2)=\left\{\varphi\in D(p) : p \varphi \in D(p) \right\}= \left\{\varphi\in H^1(\mathbb{S}_r) : \varphi' \in H^1(\mathbb{S}_r) \right\} = H^2(\mathbb{S}_r),
\end{equation}
i.e.\ the second Sobolev space of functions on the circle which are $C^1$  with square-integrable second weak derivative.
Finally, the harmonic oscillator Hamiltonian~\eqref{eq:hoHamdef} is also defined and self-adjoint on the second Sobolev space, since $D(H)= D(p^2)\cap D(q^2)= D(p^2)= H^2(\mathbb{S}_r)$. 

Let us look at its spectrum. To this end, for a generic radius $r>0$, consider the self-adjoint operator $H=H_r$ in~\eqref{eq:hoHamdef} defined on $D(H_r)=H^2(\mathbb{S}_r)$. We want to compare its spectrum to the spectrum of the Hamiltonian $H_1$ on a circle of radius 1. By making use of the unitary mapping $U:L^2(\mathbb{S}_r)\to L^2(\mathbb{S}_1)$ with $(U\varphi)(y)=\sqrt{r}\varphi(r y)$, we can map $H_r$ to the unitarily equivalent (and thus isospectral) operator
\begin{equation}
   \tilde{H}_r= U H_r U^\dag = \frac{1}{2 r^2}p^2+\frac{r^2}{2} q^2
\end{equation}
on $D(\tilde{H}_r)=UD(H_r)= H^2(\mathbb{S}_1)$.
Take a rescaled version $r^2 \tilde{H}_r$. Using perturbation theory  for small $r$, its spectrum converges to the spectrum of $p^2/2$ on the unit circle, that is $\frac{n^2}{2}$ with $n\in\mathbb{Z}$. Therefore, the spectrum of $H_r$ scales as $\frac{n^2}{2 r^2}$ for small $r$, and, in particular, it is not harmonic! 

For large $r$, $H^2(\mathbb{S}_r)$ tends to $H^2(\mathbb{R})$ and, as we will show, the spectrum converges to the usual harmonic spectrum. Notice that for arbitrary $r$ the spectrum is always pure point with finite degeneracies, since $H$ is    a bounded perturbation of the compact-resolvent operator $p^2$. 

\subsection{Eigenvalues and eigenfunctions}
The  spectrum can be obtained by explicitly constructing the eigenvalues and eigenfunctions for the problem at hand.
The harmonic oscillator's eigenvalue equation,
\begin{equation}
    -\frac{1}{2} u''(x) + \frac{1}{2} x^2 u(x) = E u(x),
    \label{eq:eigenvalue1}
\end{equation}
translates into Weber's equation 
$f''(z) -\left( \frac{1}{4} z^2 +a \right)f(z) = 0$,
for the function $f(z)=u(z/\sqrt{2})$, with $a=-E$.
A  pair of linearly independent solutions for $E\in\mathbb{R}$ and $x\in\mathbb{R}$,  having definite parity
(respectively even and odd), is given by
\begin{eqnarray}
	u_{\mathrm{e}}(E,x) &=& \mathrm{e}^{-\frac{1}{2}x^2}{}_1 F_1\left(-\frac{E}{2}+\frac{1}{4}, \frac{1}{2}; x^2\right), \nonumber\\
	u_{\mathrm{o}}(E,x) &=& x\,\mathrm{e}^{-\frac{1}{2}x^2}{}_1 F_1\left(-\frac{E}{2}+\frac{3}{4}, \frac{3}{2}; x^2\right),
    \label{eq:Weberfuns}
	\end{eqnarray}
%\end{subequations}
where ${}_1F_1(a,c; x)$ is the confluent hypergeometric function.
Therefore, the general solution of the eigenvalue equation~\eqref{eq:eigenvalue1} reads
$u(E,x) = c_{\mathrm{e}} u_{\mathrm{e}} (E,x) + c_{\mathrm{o}} u_{\mathrm{o}}(E,x)$.

Notice that $u(E,x)$ is analytic for $x\in\mathbb{R}$ and thus belongs to the second Sobolev space $H^2(-\ell,\ell)$ for any $\ell>0$.
By imposing that $u$ belong to $D(H_r)=H^2(\mathbb{S}_r)$, i.e.\ satisfy the  $C^1$ continuity conditions at the antipodal point $A$,
\begin{equation}
    u(E,\ell)=u(E,-\ell), \qquad u'(E,\ell)=u'(E,-\ell), 
\end{equation} 
with $\ell=\pi r$, one obtains the energy eigenvalues $E_n$ and the corresponding eigenfunctions $u_n$.
\begin{figure}
    \centering
    \includegraphics[width=0.6\linewidth]{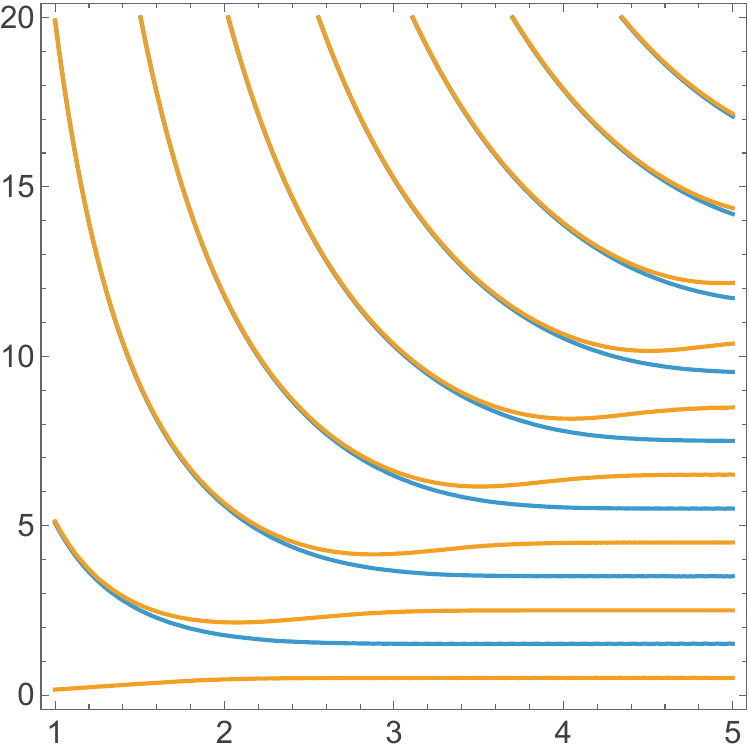}
    \caption{Even eigenvalues $E_2k$ (orange) and odd eigenvalues $E_{2k+1}$(blue) vs $\ell=\pi r$. 
    One has $E_n(\ell)\to n+\frac{1}{2}$ as $\ell\to\infty$ (spectrum of the harmonic oscillator on the line), and $E_n(\ell)\sim \lceil \frac{n}{2} \rceil^2 \frac{\pi^2}{2\ell^2}$ as $\ell\to 0$ (spectrum of a free particle on a circle).}
    \label{fig:spectrum}
\end{figure}

In particular, the even eigenfunctions have the form
\begin{equation}  
    u_{2k}(x) =  u_{\mathrm{e}}(E_{2k},x)
   ,  \qquad k\in\mathbb{N},
\end{equation}
and the corresponding energy eigenvalue $E_{2k}$ is such that $u_{2k}\in H^2(\mathbb{S}_r)\subset C^1(\mathbb{S}_r)$, that is $u_{2k}(\ell)=u_{2k}(-\ell)$ and $u'_{2k}(\ell)=u'_{2k}(-\ell)$. Since $u_{2k}$ is even (and its derivative is odd), the first condition is automatically satisfied and the second one forces the derivative to vanish at $A$, that is
\begin{equation}
    u'_{\mathrm{e}}(E_{2k},\ell) =0. 
\end{equation}
The energy eigenvalue $E_{2k}$, with $k\in\mathbb{N}$, is thus given by the $k+1$-th zero of the function $E\mapsto u'_{\mathrm{e}}(E,\ell)$.
Notice that the vanishing of the derivative of $u_{2k}$ at $A$ implies that its value cannot vanish:
\begin{equation}
\label{eq:nonvanishing}
    u_{2k}(\pi r) \neq 0,
\end{equation}
otherwise as a solution of homogeneous second order equation it must be identically zero.

Analogously, the odd eigenfunctions have the form
\begin{equation}  
    u_{2k+1}(x) =  u_{\mathrm{o}}(E_{2k+1},x)
   ,  \qquad k\in\mathbb{N},
\end{equation}
and the corresponding energy eigenvalue $E_{2k+1}$ is such that $u_{2k+1}\in H^2(\mathbb{S}_r)$. Since $u_{2k+1}$ is odd (and its derivative is even), the $C^1$ condition at $A$ implies the vanishing of the wave function there,
\begin{equation}\label{odd}
    u_{\mathrm{o}}(E_{2k+1},\ell) =0, 
\end{equation}
and the energy eigenvalue $E_{2k+1}$ is thus given by the $k+1$-th zero of the function $E\mapsto u_{\mathrm{o}}(E,\ell)$.

\begin{figure}
    \centering
    \includegraphics[width=0.85\linewidth]{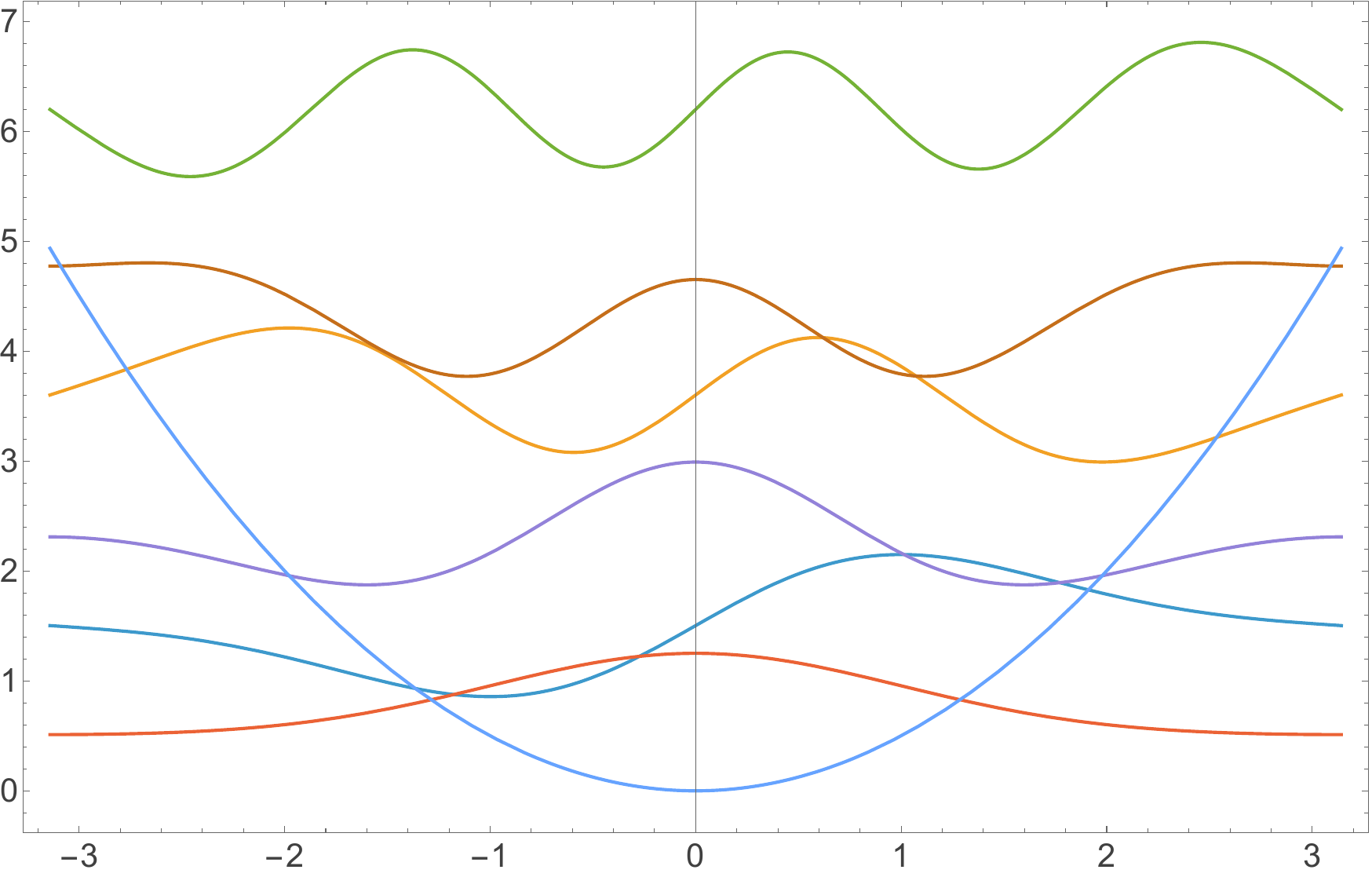}
    \caption{Eigenfunctions $u_n(x)$ vs $x$ for $r=1$ and $n=0,1,2,3,4,5$. 
    We plot the potential $x^2/2$ and set the baseline of $u_n$ to be the corresponding eigenvalue $E_n$. Notice the periodicity. For small energies we have that $u_n$ approximate the Hermite eigenfunctions of the harmonic oscillator on the line, while for large energies they approximate the trigonometric eigenfunctions of the free  particle on the circle.}
    \label{fig:eigenfunctions}
\end{figure}

The spectrum as a function of the length $\ell$ is plotted in Fig.~\ref{fig:spectrum}. The eigenfunctions are shown in Fig.~\ref{fig:eigenfunctions}.
Notice that as $\ell\to+\infty$ one gets the asymptotic behaviors
\begin{equation}
    u'_{\mathrm{e}}(E,\ell) \sim \sqrt{\pi \,\ell \,\mathrm{e}^{\ell^2}}  \frac{ \ell^{-E}}{\Gamma\left(\frac{1}{4} - \frac{E}{2}\right)},
    \qquad u_{\mathrm{o}}(E,\ell) \sim \sqrt{\frac{\pi  \,\mathrm{e}^{\ell^2}}{2 \ell}}  \frac{ \ell^{-E}}{\Gamma\left(\frac{3}{4} - \frac{E}{2}\right)}.
\end{equation}
Since the gamma function $\Gamma$ is meromorphic on $\mathbb{C}$ with simple poles at the nonpositive integers, it follows that as $\ell\to+\infty$, the zeroes of $u'_{\mathrm{e}}(E,\ell)$ tend to $\frac{1}{4}-\frac{E_{2k}}{2}=-k$, that is $E_{2k}=2k+\frac{1}{2}$, for $k\in\mathbb{N}$. Likewise the zeroes of $u_{\mathrm{o}}(E,\ell)$ tend to $\frac{3}{4}-\frac{E_{2k+1}}{2}=-k$, that is $E_{2k+1}=2k+\frac{3}{2}$. This is the spectrum of the harmonic oscillator on the line $E_n=n+\frac{1}{2}$, with $n\in\mathbb{N}$.

On the other hand, as $\ell\to 0$ one gets
\begin{equation}
    u'_{\mathrm{e}}(E,\ell) \sim  \sqrt{2E} 
    \sin(\sqrt{2 E}\, \ell),
    \qquad 
    u_{\mathrm{o}}(E,\ell) \sim   \frac{ \sin(\sqrt{2 E}\, \ell)}{\sqrt{E}},
\end{equation}
whose zeroes are respectively $E_{2k} = \frac{k^2\pi^2}{2\ell^2}$ and $E_{2k+1} = \frac{(k+1)^2\pi^2}{2\ell^2}$ for $k\in\mathbb{N}$. These are the eigenvalues
of a free particle on a circle
$E_n =  \frac{n^2\pi^2}{2\ell^2}$, with $n\in \mathbb{N}$, doubly degenerate for $n\geq 1$, as anticipated above.

 \section{Creating creation operators: Factorizing the oscillator}\label{creation}

 We now define creation and annihilation operators in analogy to the usual harmonic oscillator. Namely,

 \begin{equation}
     a=\frac{1}{\sqrt{2}}(q+ip), \qquad
     a^\dagger=\frac{1}{\sqrt{2}}(q-ip).
 \end{equation}
 Since $q$ is bounded, with $D(q)=L^2(\mathbb{S}_r)$, their domains are $D(a)=D(a^\dagger)=D(q)\cap D(p)= D(p)=H^1(\mathbb{S}_r)$. 
 
 It is easy to check that $a$ and $a^\dag$ are mutually adjoint. Indeed, for all $\psi, \varphi \in H^1(\mathbb{S}_r)$ we have
 \begin{equation}
     \langle \psi | a \varphi\rangle = \langle \psi | (q+ip) \varphi\rangle = \langle (q-ip) \psi |  \varphi\rangle = \langle a^\dag \psi | \varphi\rangle.
 \end{equation}
Moreover, exactly as on the line, they satisfy the commutation relation
\begin{equation}
   [a,a^\dag] =1 
\end{equation}
on the domain of the commutator. Indeed, for any $\psi \in D([a,a^\dag])= D(a a^\dag)\cap D(a^\dag a)$ we have that
\begin{equation}
[a,a^\dag]\psi = \frac{1}{2}\bigl((q+ip)(q-ip) \psi - (q-ip)(q+ip)\psi\bigr)=-i[q,p]\psi = \psi.
\end{equation}

Let us look closely at the domain of the commutator. One has that
  \begin{equation}
     D(a^\dag a)=\left\{\psi \in D(a): a\psi \in D(a^\dagger)\right\} = \left\{\psi \in H^1(\mathbb{S}_r): a\psi \in H^1(\mathbb{S}_r)\right\},
 \end{equation}
 that is 
 \begin{equation}
     x \psi(x) + \psi'(x) \in H^1(\mathbb{S}_r).
 \end{equation}
 Now $x$ is bounded and $C^{\infty}(\mathbb{S}_r\setminus \{A\})$  and thus $x\psi(x)$ is in $H^1(\mathbb{S}_r\setminus \{A\})$, whence $\psi'\in H^1(\mathbb{S}_r\setminus \{A\})$, that is
$\psi\in H^2(\mathbb{S}_r\setminus \{A\})$. Moreover, $H^1(\mathbb{S}_r)\subset C(\mathbb{S}_r)$ so that $x \psi(x) + \psi'(x)$ should be continuous at $A$, i.e.\ $\psi(-\ell)= \psi(\ell)$ and
\begin{equation}
    \psi'(-\ell )= \psi'(\ell) + 2 \ell   \psi(\ell),
\end{equation}
with $\ell=\pi r$.
Therefore, the domain of $a^\dag a$ is
\begin{equation}
     D(a^\dag a)=\left\{\psi \in H^1(\mathbb{S}_r)\cap H^2(\mathbb{S}_r\setminus \{A\}): \psi'(-\ell )= \psi'(\ell) + 2 \ell   \psi(\ell)\right\}.
 \end{equation}
Similarly, for the domain of $a a^\dag$, by the condition
\begin{equation}
     x \psi(x) - \psi'(x) \in H^1(\mathbb{S}_r),
 \end{equation}
 we get that
 \begin{equation}
     D( a a^\dag)=\left\{\psi \in H^1(\mathbb{S}_r)\cap H^2(\mathbb{S}_r\setminus \{A\}): \psi'(-\ell )= \psi'( \ell) - 2 \ell   \psi(\ell)\right\}.
     \label{eq:domaadag}
 \end{equation}
 Therefore,
 \begin{equation}\label{eq:equaldmin}
     D_{\mathrm{min}}:= D([a,a^\dag])= \left\{\psi \in  H^2(\mathbb{S}_r): \psi(\ell )= 0\right\} .
 \end{equation}
 
 On $D_\textrm{min}$ we have the usual factorization of the harmonic oscillator Hamiltonian:
 \begin{equation}\label{ordering}
   \frac{1}{2}\left(p^2+q^2\right)=a^\dagger a+\frac{1}{2}=aa^\dagger -\frac{1}{2}.
 \end{equation} 
 Outside of $D_\textrm{min}$ we have to be more careful. Explicitly, we have
\begin{equation}
    H=\frac{1}{2}\left(p^2+q^2\right),
    \qquad
    D(H)= H^2(\mathbb{S}_r),
\end{equation}
\begin{equation}
    H_N:=a^\dagger a+\frac{1}{2}, 
    \qquad
    D(H_N)= D( a^\dag a),
\end{equation}
\begin{equation}
    H_A:=aa^\dagger -\frac{1}{2},
    \qquad
    D(H_A)= D( a a^\dag),
\end{equation}
and because of domains they are all unequal:
\begin{equation}
    H\neq H_N\neq H_A\neq H.
\end{equation}
We will see  that this distinction is not merely formal but leads to different physics.

These three operators are all good Hamiltonians in the sense that they are self-adjoint. This was shown for $H$ in the previous section and is automatically true for $H_N$ and $H_A$ since $a^\dag a$ and $a a^\dag$ are self-adjoint, being the product of a closed operator by its adjoint~\cite{Teschl}.

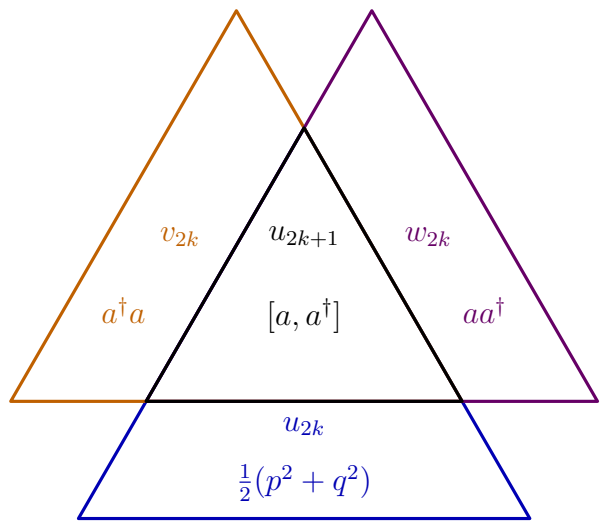
\begin{figure}[h!]
  \centering

\begin{tikzpicture}[line join=round, line cap=round, scale=1.5]

  \def\L{4}        % side length of the big triangles
  \def\O{0.70}     % overlap parameter: middle side s = O*L (0<O<1)

  \pgfmathsetmacro{\H}{sqrt(3)/2*\L}     % height of big triangles
  \pgfmathsetmacro{\s}{\O*\L}            % side length of middle triangle
  \pgfmathsetmacro{\hs}{sqrt(3)/2*\s}    % height of middle triangle
  \pgfmathsetmacro{\Y}{\H*(1-\O)}        % base height of the overlap
  \pgfmathsetmacro{\dx}{(\L-\s)/2}       % horizontal shift for red/green

  % --- Blue (centered, lower), base at y=0 ---
  \coordinate (B1) at ({-\L/2},0);
  \coordinate (B2) at ({ \L/2},0);
  \coordinate (B3) at (0,\H);

  % --- Red (upper-left), base at y=Y ---
  \coordinate (R1) at ({-\dx-\L/2},\Y);
  \coordinate (R2) at ({-\dx+\L/2},\Y);
  \coordinate (R3) at ({-\dx},{\Y+\H});

  % --- Green (upper-right), base at y=Y ---
  \coordinate (G1) at ({\dx-\L/2},\Y);
  \coordinate (G2) at ({\dx+\L/2},\Y);
  \coordinate (G3) at ({\dx},{\Y+\H});

  % --- Middle (triple overlap) equilateral ---
  \coordinate (M1) at ({-\s/2},\Y);
  \coordinate (M2) at ({ \s/2},\Y);
  \coordinate (M3) at (0,{\Y+\hs});

  % Draw triangle outlines (a bit darker)
  \draw[very thick, blue!70!black]        (B1)--(B2)--(B3)--cycle;
  \draw[very thick, orange!75!black]         (R1)--(R2)--(R3)--cycle;
  \draw[very thick, violet!80!black]       (G1)--(G2)--(G3)--cycle;
  \draw[very thick, black]               (M1)--(M2)--(M3)--cycle;

  % ---------------- Labels ----------------
  % Centers (for nice placement)
  \path (M1)--(M2)--(M3) coordinate[pos=0.333] (Mc) ; % not reliable; use centroid below
  \coordinate (Mcent) at (-0,{\Y+\hs/3-0.05});                 % centroid of middle equilateral

  \coordinate (Bcent) at (0,{\H/3});                     % centroid of blue
  \coordinate (Rcent) at ({-\dx-1},{\Y+\hs/3});             % centroid of red
  \coordinate (Gcent) at ({ \dx+1},{\Y+\hs/3});             % centroid of green

  % Middle (black)
  \node[black] at (Mcent) {$[a,a^\dagger]$};

  % Blue label: inside blue but below the black triangle
  \node[blue!70!black] at (0,{\Y/2-0.2}) {$\frac{1}{2}(p^2+q^2)$};
  \node[blue!70!black] at (0,{\Y/2+0.3}) {$u_{2k}$};
%\node[blue!70!black] at (0,{\Y/2}) {$\frac{1}{2}(p^2+q^2)\;\;\;u_{2k}$};
  % Red label: in red triangle, above overlap region
  \node[ orange!75!black] at (Rcent) {$a^\dagger a$};

  % Green label: in green triangle, above overlap region
  \node[violet!80!black] at (Gcent) {$a a^\dagger$};
  
 \node[black] at (0,2.5) {$u_{2k+1}$};
 \node[violet!80!black] at (1.1,2.5) {$w_{2k}$};
  \node[orange!75!black] at (-1.1,2.5) {$v_{2k}$};
\end{tikzpicture}

% %\end{tikzpicture}

%  }
  \caption{ Venn diagram of the domains of the Hamiltonians $H=\frac{1}{2}(p^2+q^2)$, $H_N=a^\dag a +\frac{1}{2}$ and  $H_A= aa^\dag - \frac{1}{2}$. Shown are also the eigenfunctions $u_n$, $v_n$, and $w_n$ in their respective domains. In particular, odd eigenfunctions are in the domain of the commutator and are shared by the three Hamiltonians, $u_{2k+1}=v_{2k+1}=w_{k+1}$. On the other hand, each Hamiltonian has its own even eigenfunctions ---and in particular a ground state--- characterized by a different behavior at the antipodal point $A$. }
  \label{fig:mandala}
\end{figure}

Moreover, the common domain of any pair is $D_{\mathrm{min}}$, i.e.
\begin{equation}
   D_{\mathrm{min}} = D(H)\cap D(H_N) = D(H)\cap D(H_A) = D(H_N)\cap D(H_A),
\end{equation}
see Fig.~\ref{fig:mandala}.

The three domains of $H$, $H_N$, and $H_A$ differ by the behavior of the derivative of $\psi$ at the antipodal point $A$, namely
\begin{equation}
    \psi'(A^+)= \psi'(A^-) + 2 \ell \beta \psi(A)
    \label{eq:delta}
\end{equation}
with $\beta=0,1,-1$, respectively. This is associated to a point interaction at $A$, formally arising from an additional delta potential of strength $\beta\ell$:
\begin{equation}
    V_\beta(x) = \beta \ell \delta (x-\ell), \qquad \beta = 0,1,-1.
\end{equation}
For $\beta=-1$ the singular interaction at $A$ gives rise to a bound state exponentially localized at $A$ (see section~\ref{appA} for other values of $\beta$).

Because they vanish at $A$, according to~(\ref{odd}), the odd eigenfunctions $u_{n}$  of $H$ are in $D_{\textrm{min}}$ and therefore are also the odd eigenfunctions of both $H_N$ and $H_A$ with the same eigenvalue, namely, the functions $u_n(x)= u_{\mathrm{o}}(E_{n},x)$ satisfy
\begin{equation}
    H  u_{n} = H_N  u_{n} = H_A  u_{n} = E_n u_{n}, \quad \text{for all odd } n. 
\end{equation}
See Fig.~\ref{fig:mandala}. On the other hand, according to~\eqref{eq:nonvanishing}, the even eigenfunctions of $H$ are nonvanishing at the point $A$,  so they are not in $D_{\textrm{min}}$. And since they are in $D(H)\setminus D_{\textrm{min}}$ they do not belong to $D(H_N)$ nor to  $D(H_A)$ and thus they are not shared by $H_N$ and $H_A$. In fact, the even eigenfunctions and the corresponding eigenvalues of $H_A$ and $H_N$ will be different. In order to find them, let us try to apply the usual ladder operator arguments.

\section{Climbing the ladder: algebraic argument}\label{algebraic}

Let us try to follow the standard algebraic reasoning. We have that $H_N = a^\dag a +1/2$, where the number operator $a^\dag a$ is a positive  self-adjoint operator, because $\langle\psi | a^\dag a \psi\rangle=\|a\psi\|^2\geq 0$ for all $\psi\in D(a^\dag a)$.
We have that
\begin{equation}
    a v_{\mathrm{g}} =  0, \quad  v_{\mathrm{g}}(x)=e^{-\frac{1}{2}x^2},
\end{equation}
is the ground-state eigenfunction of $H_N$ with eigenvalue $1/2$. Notice that this is  the same wavefunction of the ground state of the harmonic oscillator on the line, only truncated on $(-\ell,\ell)$. 

The next step is to apply $a^\dag$ to the ground state in order to get the first excited state. However, we immediately hit an obstruction: the wavefunction
\begin{equation}
    a^\dag v_{\mathrm{g}}(x) = 
    %\left(x- \frac{d}{dx}\right)v_0(x)= 
    \sqrt{2}x e^{-\frac{1}{2}x^2}
\end{equation}
is not in the domain of the number operator! In fact it is not even continuous at $A$, and thus we cannot apply $a^\dag$ to move further or $a$ to move back.
From this perspective this would be the end of the story, and this explains the anharmonicity of the spectrum we discussed above.

\begin{figure}
    \centering
    \includegraphics[width=0.8\linewidth]{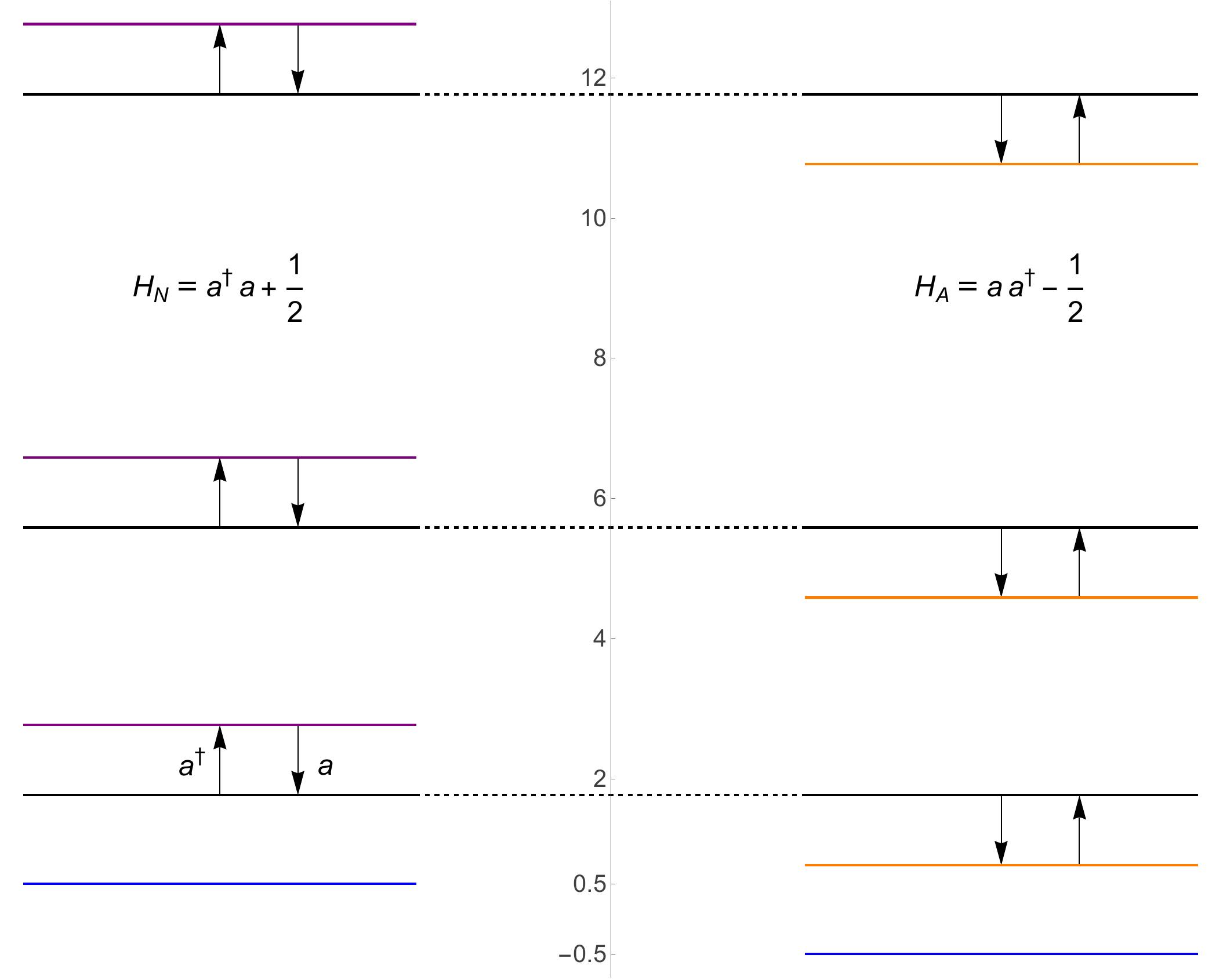}
    \caption{The lowest energy levels of $H_N$ and $H_A$ for $\ell=\pi r = 2$. Shown are the remnants of the usual ladder structure: starting on odd eigenfunctions, we obtain new eigenstates of $H_N$ (purple) by applying $a^\dag$ and of $H_A$ (orange) by applying $a$. This works only once for each odd level, which correspond to joint eigenstates of both Hamiltonians (black). Applying the opposite operator brings the state back. The blue ground states are annihilated by $a$ for $H_N$ and by $a^\dag$ for $H_A$, and have eigenvalues $1/2$ and $-1/2$ respectively.
    \label{fig:levels}}
\end{figure}

However, there is a remnant of the ladder structure if one looks at the action of $a$ and $a^\dag$ on odd eigenfunctions. Starting from $u_{2k -1}$, which is an eigenfunction of $a^\dag a$ with eigenvalue $n_{2k-1}=E_{2k-1}-1/2$ and applying the creation operator one ends up in a new eigenfunction of $a^\dag a$ and $H_N$, with eigenvalue increased by 1, as shown in Fig.~\ref{fig:levels}. Indeed,
\begin{equation}
    (a^\dag a ) a^\dag u_{2k-1} = a^\dag (a^\dag a +1 ) u_{2k-1}= (n_{2k-1}+1) a^\dag u_{2k-1} .  
\end{equation}
Here the algebraic argument can go through, because $u_{2k-1}$ is in the domain of the commutator $D_{\mathrm{min}}= D([a,a^\dag])$.

 More explicitly, one has that
\begin{equation}
    v_{2k}(x) = \sqrt{2} a^\dag u_{2k-1} (x) = x u_{2k-1}(x) - u'_{2k-1}(x),
\end{equation}
and, by using the eigenvalue equation $H_N u_{2k-1} = (n_{2k-1} +1/2) u_{2k-1}$, 
\begin{equation}
    v_{2k}'(x) = (2 n_{2k-1}+2-x^2) u_{2k-1} (x) + x u'_{2k-1}(x).
\end{equation}
Thus, one gets
\begin{equation}
    v_{2k}(-\ell)= - u'_{2k-1}(-\ell) =  - u'_{2k-1}(\ell) = v_{2k}(\ell) \neq 0 
\end{equation}
because $u'_{2k-1}$ and $u_{2k-1}$ cannot both vanish at $A$, and
\begin{equation}
    v'_{2k}(-\ell)= - v'_{2k}(\ell) = \ell v_{2k}(\ell). 
\end{equation}
Therefore, $v_{2k}$ belongs indeed to the domain of $D(H_N)=D(a^\dag a)$ in~\eqref{eq:domaadag} and is an even eigenfunction of $H_N$.
   
However, if we try to step up one more rung by applying  $a^\dag$ again we obtain the wave function
\begin{equation}
     a^\dag v_{2k}(x)= \bigl(x v_{2k}(x) - v_{2k}'(x)\bigr)/\sqrt{2},
\end{equation}
which is discontinuous at $A$ and thus not even in the domain $H^1(\mathbb{S})$ of $a$. This is exactly the same situation that we had for the ground state.  

Essentially the same story holds for $H_A = a a^\dag  -1/2$
with the role of $a$ and $a^\dag$ interchanged.
In particular, we have the ground state
\begin{equation}
    a^\dag w_{\mathrm{g}} =  0, \quad  w_{\mathrm{g}}(x)=e^{\frac{1}{2}x^2},
\end{equation}
which is remarkable as it corresponds to a negative eigenvalue of $H_A$, which equals $\frac{1}{2}(p^2+q^2)$ on $D_{\mathrm{min}}$.
Furthermore, 
\begin{equation}
    (a a^\dag) a u_{2k+1} = (n_{2k+1}-1) a u_{2k+1} .  
\end{equation}
This is again illustrated in Fig.~\ref{fig:levels}. 

The level diagram of $H_A$ also follows from more abstract arguments, since for any closed operator $B$, the product $B^\dag B$ has the same non-zero spectrum as $BB^\dag$~\cite{Teschl}. This implies that the energy levels on the right-hand side of Fig.~\ref{fig:levels} are shifted by one with respect to the left-hand side. Note that this is interesting to contrast with the full line, where the spectrum of $a^\dag a$ and $a a^\dag$ differ due to the presence of a zero eigenvalue, but where the spectrum of the Hamiltonian is independent of the ordering.

\begin{figure}
    \centering
    \includegraphics[width=0.6\linewidth]{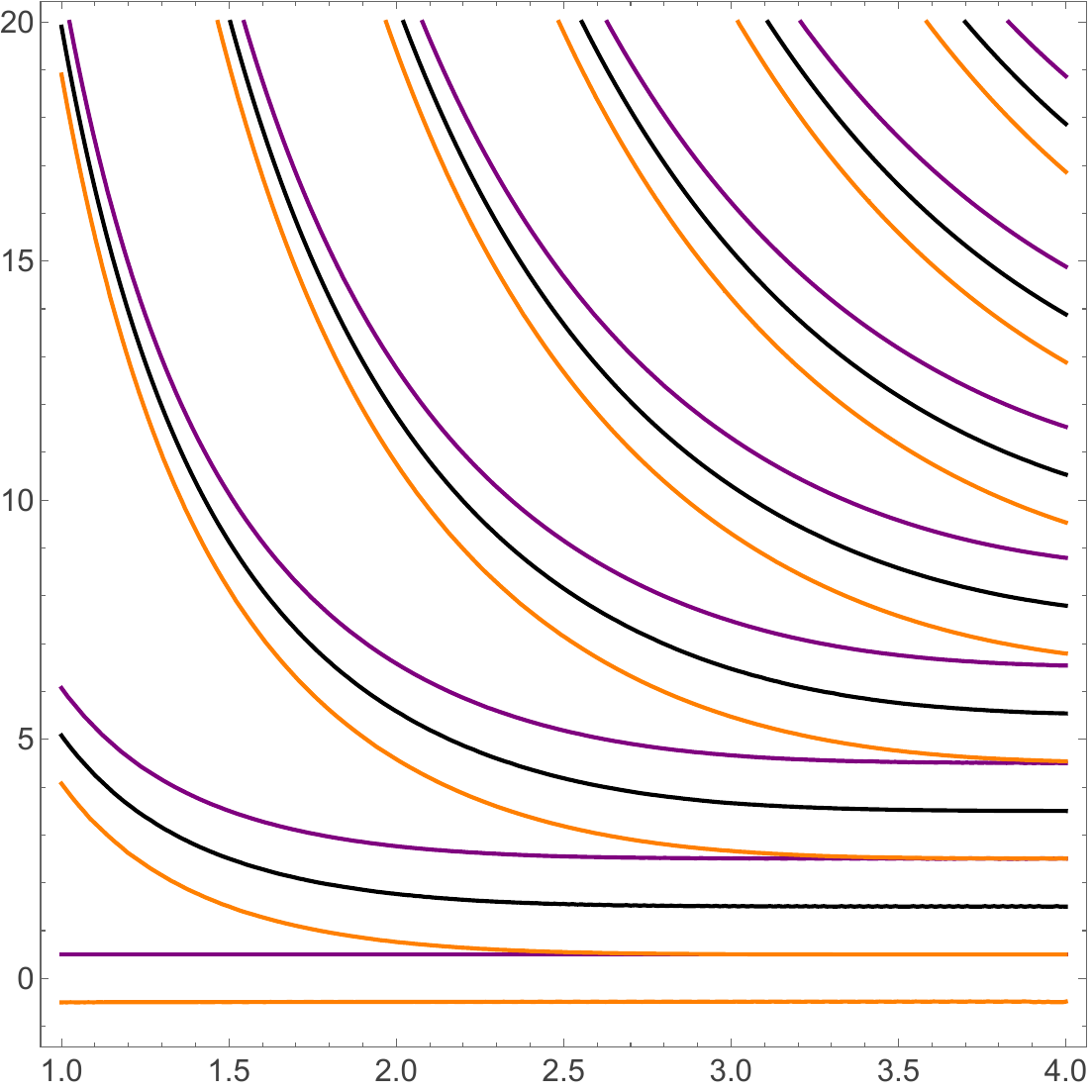}
    \caption{Even eigenvalues of $H_N$ (purple) and $H_A$ (orange) as well as the shared odd ones (black) vs $\ell=\pi r$. As $\ell\to \infty$ one recovers the spectrum of the harmonic oscillator on the line, as for $H$ in Fig.~\ref{fig:spectrum}.  Notice, however, the spectral pollution by the ground state of $H_A$ which has eigenvalue $-1/2$ independently of $\ell$.
    At variance with $H$, as $\ell\to 0$ one gets the spectrum of a free particle on a circle \emph{with} a delta potential of strength $\beta\ell$ at $A$, where $\beta=+1$ for $H_N$ and $\beta=-1$ for $H_A$. The additional point interaction lifts the degeneracy between odd and even eigenvalues by $\beta$. }
    \label{fig:comparespectra}
\end{figure}

The full spectrum of $H_N$ and $H_A$ is also easily obtained using again the Weber functions, as shown in section~\ref{appB}. We display it as a function of $\ell=\pi r$ in Fig.~\ref{fig:comparespectra}. Interestingly, while $H_N$ shows the usual alternation between even and odd parity states, $H_A$ has two even parity states as its lowest two eigenfunctions, and only then alternates.

As for the case of $H$ shown in Fig.~\ref{fig:spectrum}, the eigenvalues converge to the spectrum of the harmonic oscillator on the line as $r\to\infty$, with one notable exception: The ground energy of the antinormal-ordered Hamiltonian equals $-1/2$ for all values of $r$ and thus it survives in the limit $r\to\infty$. 

This is an instance of a curious phenomenon, named spectral pollution \cite{spectralpollution}: there are spurious eigenvalues in the limit spectrum of approximations of many physical Hamiltonians that do not belong to the true spectrum of the limit Hamiltonians. In this case one can resolve this seeming paradox by noticing that the normalized eigenfunction is exponentially localized at the antipodal point $A$ and moves away at infinity from the origin $O$. Mathematically this is expressed by its weak convergence to zero, that is to the fact that such a state becomes orthogonal to all states in the Hilbert space.

Finally, notice that, at variance with the spectrum of $H$ shown in Fig.~\ref{fig:spectrum}, as $r\to 0$ one gets the spectrum of a free particle on a circle \emph{with} a delta potential of strength $\beta\ell$ at $A$, where $\beta=+1$ for $H_N$ and $\beta=-1$ for $H_A$. The additional point interaction lifts the degeneracy between odd and even eigenvalues by $\beta$ as shown in Fig.~\ref{fig:comparespectra}.

\subsection{Spectrum of the creation and annihilation operators}

Here there is an interesting aside regarding the spectra of the creation and annihilation operators. It is immediate to see that, for any $\psi\in H^1(\mathbb{S}_r)$,
\begin{equation}
    p  e^{\frac{1}{2}x^2}\psi(x) = -i e^{\frac{1}{2}x^2} \left( x + \frac{d}{dx}\right) \psi(x) = -i \sqrt{2}e^{\frac{1}{2}x^2} a \psi(x), 
\end{equation}
that is
\begin{equation}
    a = \frac{i}{\sqrt{2}} S p S^{-1},
\end{equation}
with $S= e^{-\frac{1}{2}q^2}$ being a bounded operator with bounded inverse. 
Therefore, up to a numerical factor, $a$ is similar to the momentum operator $p$. Thus one has that the spectrum of $a$, 
\begin{equation}
    \mathrm{spec}(a) = \frac{i}{\sqrt{2}}\, \mathrm{spec}(p) =\frac{i}{\sqrt{2}r} \mathbb{Z},
\end{equation}
is pure point and purely imaginary, with eigenfunctions
\begin{equation}
    \xi_n(x) = e^{-\frac{1}{2}x^2} e^{i n \frac{x}{r}}, \qquad  n\in \mathbb{Z},
\end{equation}
and, in particular, $\xi_0=v_{\mathrm{g}}$.
This is a situation totally different from the line, where the annihilation operator has eigenvalues for all possible complex numbers (the eigenfunctions being the coherent states). Further, it is interesting to note that $ia$ is a PT-symmetric operator.

Similarly, for the creation operator one has
\begin{equation}
    a^\dag = -\frac{i}{\sqrt{2}} S^{-1} p S,
\end{equation}
whence
\begin{equation}
    \mathrm{spec}(a^\dag) = \frac{i}{\sqrt{2} r} \mathbb{Z}, \qquad \eta_n(x) = e^{\frac{1}{2}x^2} e^{i n \frac{x}{r}}, \qquad  n\in \mathbb{Z},
\end{equation}
with $\eta_0=w_{\mathrm{g}}$. We recall that on the line the spectrum of $a^\dag$ is the whole complex plane, and it has no eigenvalues.

\section{A multitude of harmonic oscillators: partial factorization and self-adjoint extensions 
%of $H_\textrm{min}$
}
\label{appA}

As we have seen above (Eq.~\ref{eq:equaldmin}), the restrictions of $H$, $H_N$ and $H_A$ to the domain of the commutator $[a,a^\dag]$,
\begin{equation}
  D_\mathrm{min}=D([a,a^\dag])=\left\{\psi \in H^2(\mathbb{S}_r): \psi(A)= 0 \right\},  
\end{equation}
are identical, even though they have different spectrum. We call it $H_{\textrm{min}}$ and it is easy to check that it is Hermitian.  Conversely, this means that $H_{\textrm{min}}$ can be extended in different ways. This is a well known subject in the theory of self-adjoint extensions~\cite{Teschl}. Here, we are interested to investigate this in the context of factorizing the Hamiltonian. 
To this end, it will be useful to introduce a more general class of creation/annihilation operators
  \begin{equation}
     a_\beta=:\frac{1}{\sqrt{2}}(\beta q+ip), \qquad 
     a_\beta^\dagger=:\frac{1}{\sqrt{2}}(\beta q-ip).
 \end{equation} with domains $D(a_\beta)=D(a^\dagger_\beta)=D(p)=H^1(\mathbb{S}_r)$. They obey 
 \begin{equation}
    [a_\beta,a_\beta^\dagger]=\beta 
 \end{equation}
 on $D([a_\beta, a_\beta^\dagger])$, 
  and lead to a partial factorization of $\frac{1}{2}(p^2+q^2)$:
 \begin{equation}
     H_\beta=a_\beta^\dagger a_\beta+\frac{1}{2}(1-\beta^2)q^2+\frac{\beta}{2},
 \end{equation} 
 on the domain
 \begin{equation}
     D_\beta=\left\{\psi \in D(a_\beta): a_\beta\psi \in D(a_\beta^\dagger)\right\}.
 \end{equation}

\begin{figure}[h!]
  \centering
  % scale the whole figure here:
  \resizebox{0.75\textwidth}{!}{%
    \begin{tikzpicture}

% ---------- parameters (in cm) ----------
\def\r{2}
\def\R{6}
\def\Hoffset{1.5}
\pgfmathsetmacro{\Hrad}{\R + \Hoffset}

\def\gap{0.2}
\def\contain{0.05}
\pgfmathsetmacro{\a}{\R - \gap}
\pgfmathsetmacro{\b}{\r + \contain}
\pgfmathsetmacro{\rlab}{\r + 0.55*(\a - \r)}

\def\dotrad{1.5pt}
\pgfmathsetmacro{\ra}{\R + 0.5}

\tikzset{>={Stealth[length=6pt,width=7pt]}}

% ==========================================================
% SHADING:
%   points lying in at least two ellipses,
%   excluding the red circle
% ==========================================================
\foreach \angA/\angB in {0/60, 0/120, 60/120}{%
  \begin{scope}
    % Clip to ellipse A (rotate the *current transform*, clip, then reset)
    \pgftransformrotate{\angA}
    \clip (0,0) ellipse [x radius=\a, y radius=\b];
    \pgftransformreset

    % Clip to ellipse B
    \pgftransformrotate{\angB}
    \clip (0,0) ellipse [x radius=\a, y radius=\b];
    \pgftransformreset

    % Subtract the red circle (even-odd rule), no rotation
    \pgfseteorule
    \clip (-100,-100) rectangle (100,100)
          (0,0) circle (1.025*\r);
    \pgfsetnonzerorule

    % Fill what's left
    \fill[gray!7] (-100,-100) rectangle (100,100);
  \end{scope}%
}

% ---------- circles ----------
\draw[line width=1pt, red]   (0,0) circle (1.025*\r);
\draw[line width=1pt, blue]  (0,0) circle (\R);
\draw[line width=1pt, black] (0,0) circle (\Hrad);

% ---------- central dot ----------
\fill[red] (0,0) circle (\dotrad);
\node[red, font=\small, right=2pt] at (0,0) {$0$};

% ---------- ellipses + labels ----------
\foreach \ang/\lbl/\vidx/\col in {
  0/{\(D_{0}\)}/0/teal!70!black,
  60/{\(D_{-1}\)}/-1/orange!85!black,
  120/{\(D_{1}\)}/1/violet!80!black}{

  \draw[line width=0.9pt, \col, rotate=\ang]
    (0,0) ellipse [x radius=\a, y radius=\b];

  \node[\col] at (\ang:\rlab) {\lbl};

  \fill[\col] (\ang+180:\rlab) circle (\dotrad);
  \node[\col, font=\small] at (\ang+180:\rlab - 0.3) {$\phi_{\vidx}$};

  \fill[\col] (\ang+180:\ra) circle (\dotrad);
  \node[\col, font=\small] at (\ang+180:\ra + 0.4)
    {$a^{\dagger}_{\vidx}\phi_{\vidx}$};

  \draw[\col, line width=0.9pt, ->, shorten >=\dotrad, shorten <=\dotrad]
    (\ang+180:\rlab) -- (\ang+180:\ra);

  \draw[\col, dashed, line width=0.8pt, ->, shorten >=\dotrad, shorten <=8pt]
    (\ang+180:\rlab-0.2) -- (0,0);
}

% ---------- region labels ----------
\node[red] at (0.4,0.4) {$D_{\mathrm{min}}$};
\node[black] at (0.2,0.8) {odd eigenfunctions};
\pgfmathsetmacro{\rout}{\R - 0.8}
\node[blue!60!black] at (210:\rout) {$D_{\mathrm{max}}$};
\node[black] at (90:\Hrad - 0.6) {$\mathcal{H}$};

\end{tikzpicture}

  }
  \caption{ Venn diagram of the domains of the Hamiltonians $H$, $H_N$ of $H_A$. Shown are also the states $\phi_\beta$ and how they behave  under their respective creation/annihilation operators. In particular, $a_\beta^\dagger \phi_\beta$ is outside of the domain of the Hamiltonians $H_\beta$, and $(a_\beta^\dagger )^2\phi_\beta$ is not even defined. 
$D_{\textrm{max}}$ denotes the domain of $H_{\textrm{min}}^\dagger$.}
  \label{fig:mandala2}
\end{figure}
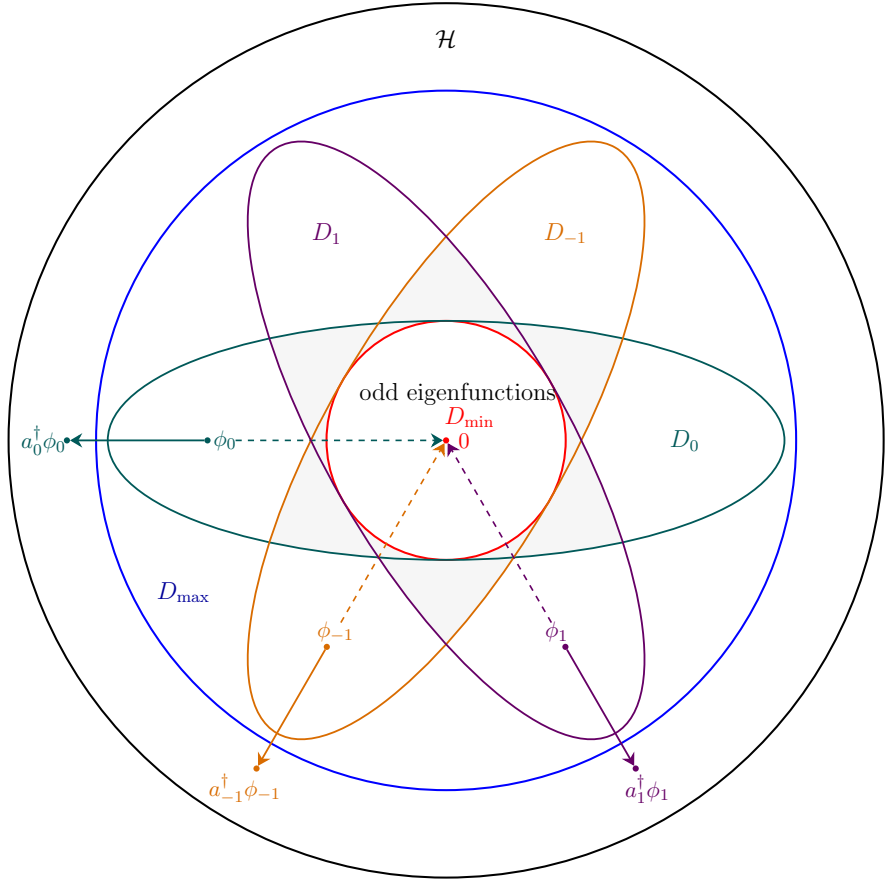

Notice that $a_{-\beta}=-a_\beta^\dagger$ and the cases $\beta=\{0,1,-1\}$ capture the natural choices of a squared momentum, of the normal ordered and of the anti-normal ordered version given in Eq. (\ref{ordering}). Since $q^2$ is bounded, the operators $H_\beta$ are self-adjoint on their respective domains, and we will now see that they characterize all self-adjoint extensions of $H_{\textrm{min}}$.

In order to understand the self-adjoint extensions of $H_\textrm{min}$, we consider the constraints we need to put on the domain of its adjoint $H_\textrm{min}^\dag$ 
\begin{equation}
  D_{\mathrm{max}}=D(H_\textrm{min}^\dag)=H^2(\mathbb{S}_r\setminus \{A\}) \cap H^1(\mathbb{S}_r)  
\end{equation}
to render it Hermitian. From the condition $\langle H^\dag_\textrm{min}u|v\rangle=\langle u|H^\dag_\textrm{min}v\rangle$ with $u,v\in D_{\mathrm{max}}$ we get 
\begin{equation}
    v(\ell)\bigl[-u'(\ell)+u'(-\ell)\bigr]^*+\bigl[v'(\ell)-v'(-\ell)\bigr]u(\ell)^*=0,
\end{equation} 
with $\ell=\pi r$.
This is solved by requiring that $u$ and $v$ satisfy the condition
\begin{equation}
    u'(-\ell)=u'(\ell)+2\ell \beta u(\ell),
    \label{eq:derivjump}
\end{equation} 
for any $\beta\in \mathbb{R}$. Therefore, 
\begin{eqnarray}
     \fl \qquad 
     D_\beta %&=&\left\{\psi \in H^2(\mathbb{S}_r\setminus \{A\}): \psi(A^+)= \psi(A^-), \; \psi'(A^+)= \psi'(A^-) - 2 \pi r \beta \psi(A) \right\}
     % \nonumber\\
     %\fl \qquad  &=& 
     =\left\{\psi \in  H^2(\mathbb{S}_r\setminus \{A\})\cap H^1(\mathbb{S}_r) \, : \, \psi'(A^+)= \psi'(A^-) + 2 \ell \beta \psi(A) \right\}.
\end{eqnarray}

A specific solution $\phi_\beta \in D_\beta$ of matching condition~\eqref{eq:derivjump} at $A$ is given by
\begin{equation}
    \phi_\beta(x) =\mathrm{e}^{-\frac{\beta}{2} x^2} ,
\end{equation}
so that the domains of the self-adjoint extensions $H_\beta$ of $H_{\mathrm{min}}$ can be written as  \begin{equation}
 D_\beta = D_\mathrm{min} \dot{+}\, \mathbb{C} \phi_\beta.   
\end{equation} 
 Indeed, any wavefunction $u\in D_\beta$ can be uniquely written as the sum $u=w+ c \phi_\beta$, with $w\in D_{\mathrm{min}}$ and $c=u(\ell)/\phi_\beta(\ell)$.

Interestingly, $\phi_\beta$ completely characterizes the self-adjoint extension $H_\beta$, since $\phi_\beta\notin D_\alpha$ for $\alpha\neq\beta$, and moreover we have that
\begin{equation}
    a_\alpha \phi_\alpha =0.
\end{equation}
See Fig. \ref{fig:mandala2} for a schematic.

The different self-adjoint extensions differ by the Robin-like matching condition~\eqref{eq:derivjump} at $A$, that can be ascribed to a point interaction produced by a delta potential at $A$.
Indeed, if one looks at the energy expectation value of a wavefunction $\psi\in D_\beta$, an integration by part gives
\begin{equation}
    \langle\psi|H_{\beta}\psi\rangle = \int dx \left(\frac{1}{2}|\psi'(x)|^2 +\frac{x^2}{2} |\psi(x)|^2 + \beta\ell \delta_\ell(x)|\psi(x)|^2 \right).
\end{equation}
The right-hand side can be extended to the full domain of the momentum $D(p)=H^1(\mathbb{S}_r)$, which is the form domain of all the self-adjoint operators $H_\beta$, that are thus characterized just by the strength of the delta potential at $A$.

The $H_\beta$ classification and the demonstration that odd states are shared across extensions while even states encode boundary physics may have implications for designing boundary-sensitive quantum systems on compact manifolds (e.g., ring-shaped traps and synthetic dimensions).

\subsection{Spectrum of the extensions} \label{appB}

With this infinite class of Hamiltonians one may consider `harmonic oscillators', let us consider their spectrum. 
Similarly to the specific case considered in Section~\ref{sec:fullcircle}, the spectrum is  obtained by explicitly constructing the spectral function for the problem at hand.

\begin{figure}
    \centering
    \includegraphics[width=0.8\linewidth]{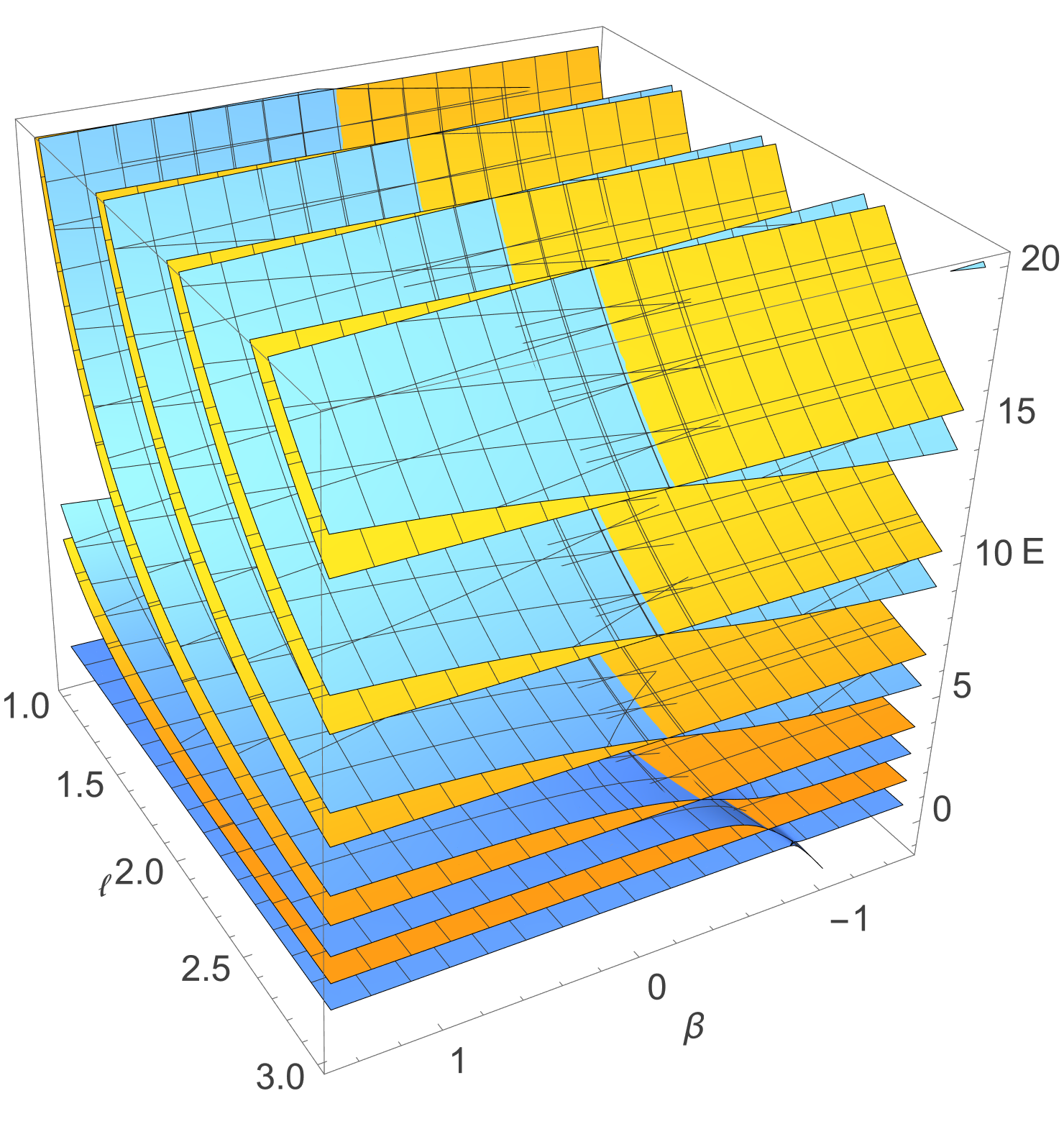}
    \caption{Low energy  spectrum of $H_\beta$ as a function of $\beta$ and $\ell=\pi r$. The yellow sheets are the odd states which do not depend on $\beta.$ For $\beta=\pm 1$ we see some integer gaps due to the remnants of the algebraic argument discussed above.}
    \label{fig:3dcontour}
\end{figure}

The harmonic oscillator's eigenvalue equation~\eqref{eq:eigenvalue1} has as independent solutions for $E\in\mathbb{R}$ and $x\in\mathbb{R}$ the even and odd Weber functions~\eqref{eq:Weberfuns}.
They are analytic for $x\in\mathbb{R}$ and thus belong to the second Sobolev space $H^2(-\ell,\ell)$ for any $\ell>0$.
By imposing that $u$ belong to $D_\beta$, i.e.\ satisfy the appropriate matching conditions at $A$, one obtain the eigenvalues $E_n^{(\beta)}(\ell)$ and the corresponding eigenfunctions $u_n^{(\beta)}$ of the Hamiltonian $H_\beta$.

The sequence of odd eigenvalues and eigenfunctions, solutions of
\begin{equation}
   u_{\mathrm{o}}(E_{2k+1},\ell)=0,  \qquad  u_{2k+1}(x) =  u_{\mathrm{o}}(E_{2k+1},x), \qquad k\in\mathbb{N},
\end{equation}
are independent of $\beta$ and shared by all the extensions $H_\beta$ and belong to the common intersection $D_{\mathrm{min}}$ of all the domains $D_\beta$.

On the other hand,  the even energy levels $E_{2k}^{(\beta)}$, solutions of
\begin{equation}
u'_{\mathrm{e}}\bigl(E_{2k}^{(\beta)},\ell\bigr) +\ell\beta u_{\mathrm{e}}\bigl(E_{2k}^{(\beta)},\ell\bigr)=0,  \qquad k\in\mathbb{N},
\end{equation}
do depend on $\beta$ and the corresponding even eigenfunctions 
\begin{equation}
u_{2k}^{(\beta)}(x) =  u_{\mathrm{e}}\bigl(E_{2k}^{(\beta)},x\bigr),  \qquad k\in\mathbb{N},
\end{equation}
belong to their respective domain $D_\beta$ only, i.e.\  $u_{2k}^{(\beta)}\notin D_\alpha$ for $\alpha\neq\beta$. 

The lower energy eigenvalues  of $H_\beta$ are shown Fig.~\ref{fig:3dcontour} as a function of $\ell=\pi r$ and of $\beta$.

\section{Conclusions}\label{conclusions}
It is remarkable that for our search of a harmonic oscillator on a circle we actually found infinitely many contestants! Out of these, the factorizing ones are $H_N$ and $H_A$, and we have shown that the standard algebraic argument only survives as a remnant of a single-step ladder. This gives a different perspective why the harmonic oscillator on the circle is not very harmonic. It provides a well-controlled counterexample to textbook ladder-operator derivations, forcing explicit attention to domains and self-adjointness. This is highly relevant to mathematical physics and to education in quantum theory. It is interesting to note that the fragility of the algebraic approach for the harmonic oscillator on the circle is qualitatively different from orbital angular momentum on compact manifolds, where the ladder approach still works, but a continous spectrum arises in addition to the usual eigenvalues~\cite{wrongangular}. Our paper could spark further studies on how algebraic structures (CCR, Weyl relations, factorizations) fragment on compact or topologically nontrivial configuration spaces, including quantum graphs and multi-point boundary defects.
\section*{Acknowledgments}
DB thanks Abe Hiroyuki for pointing out relevant references in string theory. 
PF~acknowledges support from INFN through the project ``QUANTUM'', from the Italian National Group of Mathematical Physics (GNFM-INdAM), and from the Italian funding within the ``Budget MUR - Dipartimenti di Eccellenza 2023--2027''  - Quantum Sensing and Modelling for One-Health (QuaSiModO). DB and PF acknowledge support from PNRR MUR project PE0000023-NQSTI. 

\appendix
\section{From circle to segment}\label{appC}

Wavefunctions on the circle of radius $r$ can be mapped into wavefunctions on an segment of length $2\ell = 2\pi r$ by using the Hilbert space isomorphism 
\begin{equation}
    L^2(\mathbb{S}_r)\simeq L^2(-\ell,\ell), \quad \ell=\pi r.
\end{equation}
Here, the end points $\pm\ell$ of the interval correspond to the antipodal point $A$ on the circle, and they should be thought of as the same point. 

Any functional space that requires some regularity properties are mapped through this isomorphism into the corresponding functional space on the interval, with an appropriate matching at the boundary.

In particular, the   first Sobolev space $H^1(\mathbb{S}_r)$ of square integrable functions with square integrable (weak) derivatives, which is the domain of the momentum operator on the circle, is mapped as
\begin{equation}
    H^1(\mathbb{S}_r)=\{\varphi\in C(\mathbb{S}_r) \, :\, \varphi'\in L^2(\mathbb{S}_r) \} \simeq \{\psi\in H^1 (-\ell,\ell) \, :\, \psi(-\ell)=\psi(\ell) \}.
\end{equation}

Similarly, the domain of the harmonic oscillator Hamiltonian is the second Sobolev space,
\begin{equation}
    H^2(\mathbb{S}_r)=\{\varphi\in C^1(\mathbb{S}_r) \, :\, \varphi''\in L^2(\mathbb{S}_r) \} 
\end{equation}
and is mapped into
\begin{equation}
    H^2(\mathbb{S}_r)\simeq \{\psi\in H^2 (-\ell,\ell) \, :\, \psi(-\ell)=\psi(\ell) ,\, \psi'(-\ell)=\psi'(\ell) \}.
\end{equation}

\bibliographystyle{prsty-title-hyperref}

\end{document}